\input harvmac.tex
\vskip 1.5in
\Title{\vbox{\baselineskip12pt 
\hbox to \hsize{\hfill}
\hbox to \hsize{\hfill CTP-SCU/2015017 }}}
{\vbox{
	\centerline{\hbox{A Note on GSO-Free RNS Superstrings
		}}\vskip 5pt
        \centerline{\hbox{and Pure Spinor Costraint
		}} } }
\centerline{Dimitri Polyakov$^{}$\footnote{$^{\dagger(1),(2)}$}
{polyakov@sogang.ac.kr ; polyakov@scu.edu.cn; 
twistorstring@gmail.com
}}
\medskip
\centerline{\it Center for Theoretical Physics $^{(1)}$}
\centerline{\it College of Physical Science and Technology}
\centerline{\it Sichuan University, Chengdu 610064, China}
\centerline{\it }
\centerline{\it }
\centerline{\it Institute for Information Transmission Problems (IITP)$^{(2)}$}
\centerline{\it Bolshoi Karetny per. 19/1}
\centerline{\it 127994 Moscow, Russia}
\vskip .3in

\centerline {\bf Abstract}

Using an elementary perturbative open string field theory solution
 involving a  twistor-like parameter,
we study the cohomology of new nilpotent BRST charge 
corresponding to the space-time background
defined by this solution.
 The BRST cohomology of the deformed background automatically
cuts off the GSO-odd spectrum in RNS superstring theory
and keeps the GSO-even spectrum intact, 
without  a need of GSO-projection.
The on-shell constraints in the GSO-even sector get deformed in the
 new background, corresponding
to BRST type transformations of the low-energy effective action w
ith the ghost-like commuting
spinor parameter satisfying the pure spinor constraint in $d=10$.

\Date{August 2015} 

\vfill\eject

\lref\ramond{P. Ramond, Phys. Rev. D3 (1971) 2415}
\lref\ns{A. Neveu, J. Schwarz, Nucl. Phys. B31 (1971) 86}
\lref\selfbell{D. Polyakov,  arXiv:1507.06226}
\lref\gso{F. Gliozzi, Joel Scherk, David I. Olive, Nucl.Phys.B122:253-290,
(1977)}
\lref\witsft{E. Witten, Nucl.Phys. B268 (1986) 253}
\lref\witsfts{E. Witten, Phys.Rev. D46 (1992) 5467-5473}
\lref\bars{I. Bars, Y.Matsuo,  Phys.Rev. D66 (2002) 066003}
\lref\psberkf{N. Berkovits, JHEP 0004 (2000) 018}
\lref\psberks{N. Berkovits, JHEP 09 (2004) 047}
\lref\gs{M. Green, J. Schwarz,  Phys.Lett. B109 (1982) 444-448}

\centerline{\bf  1. Introduction}

One well-known property of the Ramond-Neveu-Schwarz superstring theory
~{\ramond, \ns} is the need to implement the GSO-projection ~{\gso}, 
in order to eliminate the tachyon from the spectrum
(along with other states which, while being in the BRST cohomology, 
behave in an unnatural way in
terms of the spin-statistics properties). This , in a sense,
 raises a problem because 
the physical states of  superstring theory are most naturally 
defined in terms of the BRST cohomology
(incripting all the gauge-theoretic properties of the model), 
while the GSO-projection is essentially
an artificial operation, implemented by hands in order to exclude 
the unwanted superstring excitation modes.
Moreover, this operation breaks the BRST cohomology of the 
RNS model into two subsectors, GSO even and GSO odd which
are not completely independent or separable. 
Indeed, the on-shell operator product of 
two arbitrary GSO-even vertex operators  
doesn't contain  GSO-odd operators, 
making the GSO-even part formally ``insulate''. 
The inverse, however, is generally not true,
and this is why in RNS superstring theory, 
even with the GSO-projection implemented, the Veneziano amplitutes of all even 
states would generally get contributions from the odd sector, 
such as tachyonic poles.

On the other hand, the presence  of the GSO-odd  states in the 
BRST cohomology and in the amplitudes 
is the property of flat space-time background
 which does not necessarily extend to other
target space backgrounds with different cohomologies.
One particularly convenient and efficient tools
 to explore the BRST cohomologies in various backgrounds
is string field theory with the equations of motion
{\witsft, \witsfts}:
\eqn\grav{\eqalign{Q\Psi+\Psi\star\Psi=0}}
 which is background-independent by construction.
That is, assume
$\Psi_0$ is a solution
of the equation (1). Then the form of (1) is invariant under the shift
\eqn\lowen{
\Psi\rightarrow{\tilde{\Psi}}=\Psi+\Psi_0}
with the simultaneous shift of the BRST charge
$Q\rightarrow{\tilde{Q}}$ , so that $Q^2={\tilde{Q}}^2=0$ and
the new nilpotent charge ${\tilde{Q}}$ defined according to
\eqn\lowen{{\tilde{Q}}\Psi=Q\Psi+\Psi_0\star\Psi+\Psi\star\Psi_0}
for any $\Psi$.
Then the new BRST charge ${\tilde{Q}}$ defines the new cohomology, different
from that of the original charge $Q$, corresponding to string theory in a
 new background, depending on the structure of $\Psi_0$. The advantage of 
this approach is that, in principle, it allows to
 explore the string theory in  new geometrical backgrounds
(e.g. in a curved geometry, such as AdS) while technically using
the operator products of the old string theory 
(say,  in originally flat background)
for the vertex operatos
in the new BRST cohomology, defined by ${\tilde{Q}}$.
In this letter we consider an example of an elementary open SFT solution
describing simple perturbative background deformation and eliminating the odd sector
from the cohomology. The deformation is 
parametrized by a commuting spinor parameter
which has to satisfy the pure spinor constraints in order to ensure the 
nilpotence of the BRST charge. The deformation of the cohomology results in
eliminating the GSO-odd operators from the spectrum and in some  modifications
of the on-shell constraints for the space-time fields multiplied by the appropriate
vertex operators in the GSO-even sector. In terms of the low-energy effective action in space-time, 
this particularly corresponds to deforming the space-time fermions by the ghost-type
bosonic pure spinor variables (BRST-type transformation).
In the rest of the note, we shall describe this background deformation
and analyze the cohomology change, leading
to the GSO-free superstring with the modified even sector.

\centerline{\bf 2. Elementary OSFT Solution and Cohomology Change}

In our recent previous work ~{\selfbell}
we particularly pointed out a class of the elementary OSFT solutions
describing perturbative background deformations.
Namely, Let $V_{i}(z,p) (i=1,...)$ be the set of all physical vertex operators
in string theory in the cohomology of the original BRST charge $Q$
(primary fields of ghost number 1 and conformal dimension 0)
and $\lambda^i(p)$ are the corresponding space-time fields
(where $p$ is the momentum in space-time and we suppress the space-time 
indices for the brevity).
Then the string field
\eqn\grav{\eqalign{\Psi_0=\sum_i\lambda^iV_i}}
is the solution of (1) provided that the zero $\beta$-function
conditions:
\eqn\grav{\eqalign{\beta_{\lambda^i}=0}}  are imposed on the space-time fields 
in the leading order of the perturbation theory.
The related BRST charge deformation is then 
\eqn\lowen{Q\rightarrow{\tilde{Q}}=Q+\sum{\lambda_i{V^i}}}
with ${\tilde{Q}}$ acting on string fields according to the prescription (3),
which details we shall also discuss below.
One particularly simple OSFT solution, relevant to our discussion is
\eqn\grav{\eqalign{\Psi_0={\xi_\alpha}ce^{-{1\over2}\phi}\Sigma^\alpha
+B_mce^{-\phi}\psi^m}}
where $\phi$ is the bosonized superconformal ghost field,
$\Sigma^\alpha$ ($\alpha=1,...,16$) is the 16-component
spin operator for the matter RNS fermions and
$\xi_\alpha,B_m$ are some parameters which, for simplicity, are
considered constant in this work (unless specified otherwise).
The 16-component spin operator is defined in the standard way.
That is, take 10 RNS fermions $\psi_m(m=1,...,10)$, use them to make
5 complex fermions according to
$\Lambda_1=\psi_1+i\psi_2,...,\Lambda_5=\psi_9+i\psi_{10}$
and bosonize as
\eqn\grav{\eqalign{\Lambda_i=e^{i\varphi_i}\cr
i=1,...,5}}
Define the $32$-component spin operator according to
\eqn\grav{\eqalign{{{S}}_\alpha=
exp{\lbrace}{{1\over2}\sum_{i=1}^5{\pm}{i\varphi_i}}{\rbrace}
}}
with $\alpha=1,...,32$ corresponding any of 32 possible $\pm$-combinations.
This 32-component, operator can, in turn, be decomposed into
two 16-component operators $S=\Sigma\oplus{\tilde{\Sigma}}$
with $\Sigma_\alpha(\alpha=1,...,16)$  corresponding
to 16 $\pm$-configurations with even number of $+$'s and
${\tilde{\Sigma}}_\alpha$ corresponding to 16 $\pm$-configurations
with odd number of $+$'s. Such a  decomposition can, of course, be viewed 
as a split of the spin operator $S$ into GSO-even and odd components,
however, note that the string field $\Psi_0$ of (7) is the solution
of the OSFT equations of motion, invariant under the change
$\Sigma\rightarrow{\tilde{\Sigma}}$, 
bearing no reference to the GSO-projection. For the sake of certainty
 we  shall concentrate on the SFT solution (7)
 involving the spin operator $\Sigma$ with the even number
of the $+$-components. Furthermore, for our purposes we shall mostly ignore
the spin 1 component of the solution, concentrating on the spin ${1\over2}$.
First of all, we shall specify the related deformation of the BRST charge.
Again, for our purposes it is sufficient to limit ourselves to string 
fields constrained
to combinations of the on-shell operators, i.e. the dimension 0 primaries.
Using (9), it is straightforward to check that
the nilpotence of the new BRST charge:
\eqn\grav{\eqalign{{\tilde{Q}}=Q+{\xi_\alpha}ce^{-{1\over2}\phi}\Sigma^\alpha}}
implying
\eqn\grav{\eqalign{{\tilde{Q}}\Psi=Q\Psi+\xi_\alpha(ce^{-{1\over2}\phi}\Sigma^\alpha\star\Psi
-\Psi\star{ce^{-{1\over2}\phi}\Sigma^\alpha})}}
requires the constraint
\eqn\grav{\eqalign{
\xi_\alpha\xi_\beta\lbrack
ce^{-{1\over2}\phi}\Sigma^\alpha{\star}ce^{-{1\over2}\phi}\Sigma^\beta\star\Psi\rbrack
=0
}}
for some string field $\Psi$.
This constraint requires the vanishing of the worldsheet correlators
\eqn\grav{\eqalign{
\xi_\alpha\xi_\beta
<<\Psi, ce^{-{1\over2}\phi}\Sigma^\alpha{\star}ce^{-{1\over2}\phi}\Sigma^\beta\star\Psi>>
\cr
=\xi_\alpha\xi_\beta
<<h{\circ}f_1^3\circ\Psi(0)h{\circ}f_2^3\circ(ce^{-{1\over2}\phi}\Sigma^\alpha)(0)
h{\circ}f_3^3\circ(ce^{-{1\over2}\phi}\Sigma^\beta)(0)>>}}
where
the conformal transformations
\eqn\grav{\eqalign{
f_k^3(z)=e^{{2i\pi(k-1)}\over{3}}({{1-iz}\over{1+iz}})^{2\over{3}}\cr
k=1,2,3
}}
 map the string fields living on three separate worldsheets
to three wedges of a single disc,
while the transformation
$${h(z)=i{{1-z}\over{1+z}}}$$
further maps this disc to the halfplane (e.g. see ~{\bars} for explanations
on the star product details in SFT).
Although the action 
of the  conformal transformations (14)  is complicated
for generic $\Psi$, it acts trivially on dimension 0 on-shell primaries
in the three-point correlator,
simply shifting them from $0$ to the points ${\sqrt{3}},0,-{\sqrt{3}}$ 
respectively. Therefore the value of this correlator is given simply
by the structure constant in front of the $(z-w)^0$-term in the operator
prodoct of
$\xi_\alpha{c}e^{-{1\over2}\phi}\Sigma^\alpha(z)$
with itself at $w$.
The operator product is easily evaluated as
\eqn\grav{\eqalign{
\xi_\alpha{c}e^{-{1\over2}\phi}\Sigma^\alpha(z)\xi_\beta{c}e^{-{1\over2}\phi}\Sigma^\beta(w)
\sim
(z-w)^0\xi^\alpha\gamma^m_{\alpha\beta}\xi^\beta
\partial{c}c{e^{-\phi}}\psi_m + O(z-w)}}
so the vanishing of the star product (13) to ensure the nilpotence
of ${\tilde{Q}}$ requires
\eqn\grav{\eqalign{
\xi\gamma^m\xi=0}}
i.e. the pure spinor constraint  on $\xi$.
Note that  the similar
 constraint could have been obtained
by requiring the nilpotence of the charge:
\eqn\grav{\eqalign{
Q+\xi_\alpha\oint{dz}e^{-{1\over2}\phi}\Sigma^\alpha}}
with the second term in (17) being structually reminiscent
the BRST sharge $\sim\oint\lambda^\alpha{d_\alpha}$ with the Green-Schwarz
 variable $d_\alpha$ corresponding to the space-time supercurrent
at picture $-{1\over2}$
There is some subtlety here.
The operator
$\oint{dz}{e^{-{1\over2}\phi}}\Sigma^\alpha(z)$
is known to be the operator of the space-time charge. 
This operator is an anticommuting space-time spinor. The anticommutation
property can be easily seen
from analyzing the midpoint OPE of the integrand 
with itself which only contains the odd powers of $z-w$.
On the other hand, the corresponding unintegrated operator,
$c{e^{-{1\over2}\phi}}\Sigma^\alpha(z)$, is 
a $commuting$ space-time spinor since its midpoint OPE only
contains the even powers of $z-w$. 
(note that the unintegrated operator is, strictly speaking, $not$ related
to the space-time supercharge, as its OPE with itself
does not contain a momentum generator).
In the on-shell string theory 
this spin-statistics change is not of significance, since
 the unintegrated and integrated vertices
are related by the $b-c$ picture-changing transformation, 
defined by the nonlocal BRST-invariant operator
$Z=b\delta(T)$ where $T$ is the full matter$+$ghost
 worldsheet stress-energy tensor.
The $Z$-operator , particularly mapping 
unintegrated on-shell vertex operators into integrated ones,
is the $b-c$ analogue of the usual picture-changing operator 
for the $\beta-\gamma$ system and, being an
anticommuting fermion
of conformal dimension $0$, twists the spin-statistics, 
when applied to the on-shell operators.
In string field theory, however, the $Z$-transformation becomes 
ill-defined and it is essential that
the star product operates with the unintegrated vertices only. 
For this reason, it is essential
 that the $\xi_\alpha$-parameter in the deformed BRST charge (10) 
is a $commuting$ space-time spinor, i.e. a
twistor-like parameter. The nilpotence of the BRST charge thus entails 
the pure spinor constraint (17)
on this parameter.
 The essential property of the deformation (10) of the BRST charge is that
it eliminates all the GSO-odd modes from the spectrum since the 
OPE of the deformation
term with any GSO-odd operator contains a branch point. For example, for
the tachyon one gets
\eqn\grav{\eqalign{
:c{e^{-{1\over2}\phi}}\Sigma^\alpha:(z):(p\psi)e^{ipX}:(w)
\sim{\sqrt{z-w}}
{e^{-{1\over2}\phi}(\gamma^mp_m)_{\alpha\beta}\Sigma^\beta{e^{ipX}}(w)
}}}
and similarly for any other GSO-odd operator.
For the GSO-even sector, the deformation preserves analiticity,
however, the on-shell BRST-invariance constraints are modified
in the new cohomology. 
Consider the operator for the supermultiplet including the photon:
\eqn\grav{\eqalign{V_{ph}=A_m(p)\times\lbrace
c(\partial{X^m}+i(p\psi)\psi^m)+{1\over2}\gamma\psi^m\rbrace{e^{ipX}}
\cr+u_\alpha(p)ce^{-{1\over2}\phi}\Sigma^\alpha{e^{ipX}}
}}
where $A_m(p)$ is the photon's polarization vector and $u_\alpha(p)$
is the commuting space-time spinor and calculate its commutator with 
the modified BRST charge (10). 
Note that, although this operator appears to have an uncertain spin-statistics,
with the first term being odd and the second even, this uncertainty 
is merely the artefact of our $b-c$ picture choice (with the both of the operators being at unintegrated picture). Indeed, all the NS operators are even
at the integrated picture  and odd at the unintegrated one, 
while the Ramond operators are odd at the integrated picture
and even at the unintegrated one. Therefore the spin-statistics of the operator
(19) can be made certain (even) by a picture-changing transformation,
  bringing the first term to the integrated picture, and leaving the
second unintegrated. Changing the $b-c$ pictures of the operators
also entailes switching from anticommutators to commutators
(or vice versa) in the commutation relations with the BRST charge.
In particular, with the picture choice in (19) this implies the
anticommutation of $Q$ with the first term and the commutation with the second.

Straightforward calculation using (10), (11) gives:
\eqn\grav{\eqalign{\lbrace{\tilde{Q}},V_{ph}(p)\rbrack
=\partial{c}ce^{-\phi}\psi_m{e^{ipX}}(p^2A^m(p)+\xi\gamma^m{u})
+\partial{c}ce^{-{1\over2}\phi}{e^{ipX}}
\Sigma^\alpha(p^2u_\alpha+\xi^\beta\gamma^{mn}_{\alpha\beta}F_{mn})
\cr
+{i\over2}ce^{{1\over2}\phi-\chi}\Sigma^\alpha{e^{ipX}}
\gamma^m_{\alpha\beta}(p_mu^\beta(p)+{\xi^\beta}A_m)
-{i\over2}ce^{i\chi}e^{ipX}(pA(p))}}
with $F_{mn}=p_{\lbrack{m}}A_{n\rbrack}(p)$ and
the BRST-invariance imposing the on-shell constraints:
\eqn\grav{\eqalign{p^2A^m(p)+\xi\gamma^m{u}=0\cr
\gamma^m_{\alpha\beta}(p_mu^\beta(p)+{\xi^\beta}A_m)=0}}
with the Lorenz gauge condition
\eqn\grav{\eqalign{
pA(p)=0}}
Here the $\lbrace\rbrack$ symbol stands for the anticommutator of the BRST
with the first term and the commutator with the second, as explained above.
Note that the $\lbrace\lbrack$-operation ensures that the on-shell constraints
(20) have the definite (even) spin-statistics.

The remaining constraint:

\eqn\lowen{
p^2u_\alpha+\xi^\beta\gamma^{mn}_{\alpha\beta}F_{mn}=0}

 simply follows
from the second equation in (21) modulo the gauge condition).
The equations (21), (22) are equivalent to the the Euler-Lagrange
equations for the low-energy effective action in the position space:
\eqn\grav{\eqalign{
S=\int{d^{10}X}{\lbrace}{1\over4}F_{mn}F^{mn}+
\gamma^m_{\alpha\beta}(u^\alpha\partial_mu^\beta(x)
+{\xi^\alpha}D_mu^\beta){\rbrace}\cr
D_m=\partial_m+A_m
 }}

If $\xi$ is a pure spinor parameter (also satisfying the on-shell Dirac
constraint), the action (24) is gauge-invariant and can be obtained from the 
standard QED by the BRST-type transformation 
$u_\alpha\rightarrow{u_\alpha+\xi_\alpha}$
with $\xi_\alpha$ playing the role of the ghost-like variable.
Note that this is formally reminiscent of the BRST transformation
$\delta\theta_\alpha=\xi_\alpha$ in the pure spinor formalism ~{\psberkf, \psberks}
where $\theta$ is the Green-Schwarz variable ~{\gs}.
The cohomology of ${\tilde{Q}}$ is thus GSO-projected by construction,
just like in Green-Schwarz or pure spinor formalisms.
However, the cohomology change due to the  deformation of the BRST charge in the 
RNS formalism is that Ramond and NS sectors  are no longer separable. 
That is, in order
to obtain a physically meaningful low-energy limit, in the new cohomology
the  physical operators always have to combine NS and Ramond pieces.
For example, if one considers a photon operator 
$\sim{\lbrace}c(\partial{X^m}+i(p\psi)\psi^m)
+{1\over2}\gamma\psi^m\rbrace{e^{ipX}}$ of the old cohomology,
without adding the Ramond piece, its  commutator with
${\tilde{Q}}$ would lead to extra constraint $F_{mn}=0$ which would
ruin all the physical content. In other words, in the new cohomology
the physical operators have a form,
 manifestly consistent with the space-time supersymmetry.

\centerline{\bf Conclusion}

In this note we have considered the deformation of the BRST charge
by the Ramond current, consistent
with elementary solution in open superstring field theory.
The deformation is parametrized by the commuting  spinor $\xi_\alpha$
(which, for simplicity, we considered constant in this work, however, it is in
 general sufficient that it satisfies the Dirac on-shell constraint). 
 The nilpotence of the new BRST charge imposes the pure spinor
constraint on the deformation parameter, and the  cohomology of the new BRST 
charge automatically cuts off the GSO-odd sector of  superstring theory,
effectively making the RNS formalism GSO-free, on equal footing
with Green-Schwarz or pure spinor formalisms.
In the new cohomology, NS and Ramond sectors essentially get intertwined,
with the physical operators mixing the NS and Ramond ingredients
 (unlike the original BRST cohomology where these ingredients can be separated).
It is known that the GSO-projections, while formally eliminating the tachyon
and other odd states from the spectrum, still cannot eliminate the tachyonic
poles and the poles from other odd states from the amplitudes.
It is natural to expect that the scattering
amplitudes of the vertex operators from the new cohomology should have
the tachyonic and GSO-odd related poles eliminated from their structure.
Roughly speaking, the cancellation should occur as a result of the contributions
from the NS and Ramoind ingredients of the operators.
In the present note, we basically didn't concentrate on the mutual
normalizations of the NS and Ramond parts, choosing it arbitrarily
by hands in (19). Such a normalization, however, must be significant
for the cancellation of the poles from the GSO-odd sector and
must be fixed in the process of calculating some concrete
scattering amplitudes.
This shouldn't be a hard calculation and we hope to demonstrate it soon 
explicitly in the future work.

\listrefs

\end